\documentclass[noshowpacs, showkeys, amsmath, amssymb, aps, prb, floatfix, reprint]{revtex4-1}

\usepackage{graphicx,xcolor}
\usepackage{pgfplots}
\usetikzlibrary{calc}

\renewcommand{\figurename}{Figure}

\newcommand{\suppfig}[1]{Supplementary Figure~S~-~#1}
\newcommand{\suppnote}[1]{Supplementary Note~#1}

\newcounter{suppEq}
\newenvironment{suppEq}{\refstepcounter{suppEq}\equation}{\tag{S-\thesuppEq}\endequation}

\newcommand{\adde}[1]{#1}

\newcommand{\romaUNI}{Dipartimento di Fisica - Sapienza Universit\`{a} di Roma, P.le Aldo Moro 5, I-00185 Roma, Italy}
\newcommand{\milanoCNR}{Istituto di Fotonica e Nanotecnologie - Consiglio Nazionale delle Ricerche (IFN-CNR), P.za Leonardo da Vinci, 32, I-20133 Milano, Italy}
\newcommand{\milanoPOLI}{Dipartimento di Fisica - Politecnico di Milano, P.za Leonardo da Vinci, 32, I-20133 Milano, Italy}

\begin{document}

\title{Thermally-Reconfigurable Quantum Photonic Circuits at Telecom Wavelength\\by Femtosecond Laser Micromachining}

\author{Fulvio Flamini}
\author{Lorenzo Magrini}
\author{Adil S. Rab}
\author{Nicol\`o Spagnolo}
\author{Vincenzo D'Ambrosio}
\author{Paolo Mataloni}
\author{Fabio Sciarrino}
\email{fabio.sciarrino@uniroma1.it}
\affiliation{\romaUNI}

\author{Tommaso Zandrini}
\author{Andrea Crespi}
\author{Roberta Ramponi}
\author{Roberto Osellame}
\email{roberto.osellame@polimi.it}
\affiliation{\milanoCNR}
\affiliation{\milanoPOLI}

\begin{abstract}
The importance of integrated quantum photonics in the telecom band resides on the possibility of interfacing with the optical network infrastructure developed for classical communications. In this framework, femtosecond laser written {integrated photonic} circuits, already assessed for quantum information experiments in the 800~nm wavelength range,  have great potentials. In fact these circuits, written in glass, can be perfectly mode-matched at telecom wavelength to the in/out coupling fibers, which is a key requirement for a low-loss processing node in future quantum optical networks. In addition, for several applications quantum photonic devices will also need to be dynamically reconfigurable. Here we \adde{experimentally} demonstrate the high performance of femtosecond laser written photonic circuits for quantum experiments in the telecom band and we \adde{show the} use of thermal shifters, also fabricated by the same femtosecond laser, to accurately tune \adde{them. 
State-of-the-art manipulation of single and two-photon states is demonstrated, \adde{with fringe visibilities greater than 95\%. This opens 
 the way to the realization of reconfigurable quantum photonic circuits on this technological platform}.}
\end{abstract}


\maketitle

\section{Introduction}

The inherent advantages of using photons to encode information in computing applications \cite{OBri07,Walt05,Lafl01,Walm05} and to implement future quantum networks \cite{Duan01,Kimb08} have drawn in the last years increasing attention to the field of quantum optics. In particular, a significant boost has been recently provided by the use of integrated photonic circuits as a stable and compact platform for producing quantum optical devices \cite{Poli08,Mars09,Smit09,Tanz12,Bonne12}. Among the different fabrication technologies employed to produce integrated optical circuits, femtosecond laser writing is rapidly gaining a relevant role thanks to its several advantages \cite{Mars09,Cres13,Szam14}. In fact this technique, being maskless, allows one to rapidly prototype high-quality photonic circuits in glass with a very fast optimization loop and cost-effective process. In addition, it produces low birefringence waveguides, which are required for polarization encoding of single photons \cite{Sans12, Heilmann14, Corrielli14}, and has unique three-dimensional capabilities, enabling unprecedented device layouts \cite{Meany12,Spagnolo13,Poulios14}. Employing this technology for devices working in the telecom band can add further advantages since the insertion losses of these devices in an optical fiber network are very low, thanks to the almost perfect mode matching and to the reduced propagation losses \cite{Arriola13,Meany14}. The quality of photonic circuits written by femtosecond lasers has been proven\adde{ by the demonstration of several classical devices in glass as waveguide amplifiers, lasers, broadband couplers and demultiplexers \cite{DellaValle05, Osell08, Chen08, Eaton09}, as well as electro-optic modulators in crystals. \cite{Liao08, Horn12}} Although the use of this fabrication technology has been widely employed in the production of quantum devices in the 800~nm wavelength range, no investigation on the performances of these devices at telecom wavelength in the quantum regime has been performed yet.

For future application of integrated photonic devices to quantum information, a very important feature is their reconfigurability \cite{Matt09,Smit09,Shad12,Bonne12,Metc14,Bonn12,Humph14,Zhang14}. This capability could bring a two-fold advantage: on the one hand it would prove very useful to compensate for errors due to fabrication tolerances; on the other hand, it would allow one to implement several quantum protocols that require changing the functionality of the circuit dynamically. In the first case, a static compensation is necessary; in the second case, the needed device reconfiguration rate is rather low due to the currently limited brightness of multiphoton sources that require long measurement times. For this reason, thermal tuning is considered a convenient choice due to its simple implementation and its stability. In fact thermal shifters are widely employed in several integrated photonic platforms, from silicon photonics to silica-on-silicon \cite{Matt09,Smit09,Shad12,Bonne12,Metc14}; however, they have never been employed in femtosecond laser written circuits.

In this work we report on the fabrication, characterization and application of a reconfigurable integrated photonic circuit for quantum information at telecom wavelengths. Femtosecond laser micromachining is employed to inscribe waveguide Mach-Zehnder interferometers in a glass chip and to pattern gold resistive heaters on the chip surface. A reliable modeling of the thermal tuning is reported, and a full experimental characterization of the response of the interferometers is performed by injecting coherent light. Moreover, tunable Hong-Ou-Mandel interference and super-resolved fringes based on N00N states are successfully demonstrated with two-photon input states.

\section{Materials and methods}

\subsection{Fabrication of the integrated device}

\begin{figure}[t]
\centering
\includegraphics{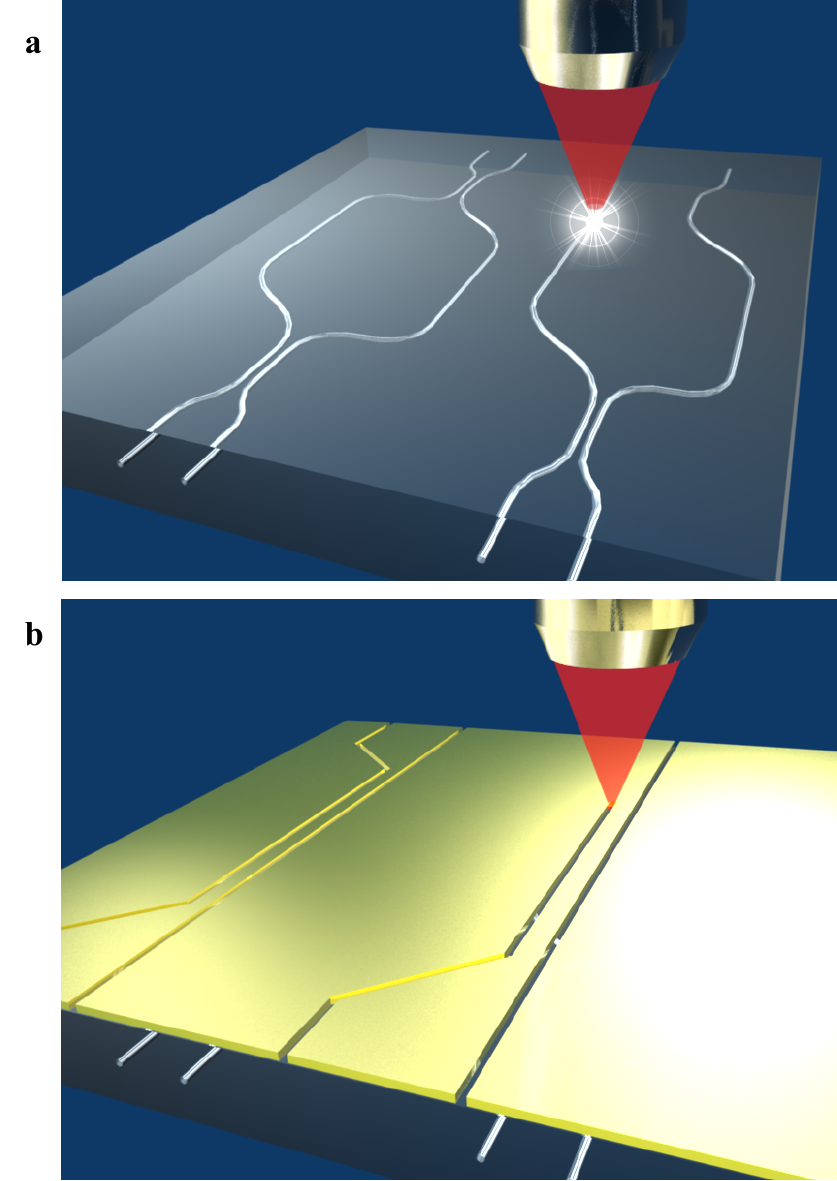}
\caption{Femtosecond laser microfabrication process. ({\bf a}) Direct writing of Mach-Zehnder interferometers in the bulk of a borosilicate slide. (\textbf{b}) After gold coating of the sample top surface, the resistors are patterned by ablation with the same femtosecond laser. Accurate alignment ($\sim$ 1~$\mu$m) between the resistors and the Mach-Zehnder arms is achieved by reference markers inscribed on the glass surface together with the interferometers.}
\label{fig:fs_fab}
\end{figure}

The fabrication process of the reconfigurable devices consists in three steps: fabrication of the three-dimensional waveguide circuits by femtosecond laser writing, deposition of a gold layer on the surface of the sample, patterning of the surface metallic layer (again realized with femtosecond laser pulses) to define the  resistor shape (see Fig.~\ref{fig:fs_fab}).
Waveguides are fabricated in Corning {EAGLE2000} alumino-borosilicate glass substrate using a Yb:KYW cavity-dumped mode-locked laser oscillator, that produces pulses with 1030~nm wavelength, 300~fs duration and 1~MHz repetition rate. To inscribe the waveguides, pulses with 330~nJ energy were used, focused by a 50$\times$ objective (NA~=~0.6), with a translation speed of 20~mm~s$^{-1}$ (three-axes translation stage, Aerotech ANT). 
The fabricated waveguides yield single mode operation at 1550~nm, with mode diameter ($1/e^2$) of about 13.5~$\mu$m and propagation losses \adde{of about} 0.6~dB~cm$^{-1}$. Such mode size enables coupling losses to standard single-mode telecom fibres \adde{of 0.2~dB} per facet (retrieved from the overlap integral of the measured mode-intensity profiles for waveguides and fibres \adde{\cite{Osel04}}).

The interferometers are composed of two cascaded directional couplers, inscribed at 70~$\mu$m depth below the substrate surface, in which the two waveguides are brought as close as 14~$\mu$m for a length of 200~$\mu$m to achieve a balanced splitting ratio. In between the two couplers, the waveguides are straight and separated by 1.02~mm for a 12-mm long segment. Such straight segments have different depth in the two arms: one is brought up to 25~$\mu$m depth, to be closer to the surface where resistors are placed, while the other is brought down to 115~$\mu$m depth. Curved waveguide segments consist in sinusoidal bends, with a minimum curvature radius of 60~mm. The overall length of the chip is 45~mm\adde{; thickness is 1~mm}.
This geometry for the interferometers, which presents two quite long and well separated straight segments in the central part, was chosen to facilitate the  characterization of thermal diffusion effects and model the operation of the thermal shifter, described in Sec.~\ref{sec:invThermal}. Much more compact devices could be realized by choosing a smaller separation between the waveguides in the central part and by minimizing the length of the straight segments.
Ablation lines were also machined on the substrate surface just after the waveguide inscription, to serve as alignment markers for the following processes.

A 50-nm gold layer was deposited on the substrate surface using a sputter coater (Cressington 108auto).
The  resistor pattern was defined by laser ablation, focusing on the sample surface the same femtosecond laser beam used for waveguide inscription. Here 100~nJ pulses were employed, with a translation speed of 1~mm~s$^{-1}$. Each laser scan produces a 2.7~$\mu$m wide ablation line in the gold layer, with negligible damage of the glass surface. To safely isolate each  resistor from the neighbouring ones, nine successive scans, spaced by 0.5~$\mu$m, were performed. Six rectangular resistors (0.3~mm wide and 20~mm long) were inscribed on the sample surface, terminating with larger areas employed to facilitate the electric connection to external circuits. Each resistor is placed just above one interferometer's arm. Some of the fabricated interferometers have two resistors, placed respectively above each of the two arms; other interferometers have been inscribed at different distances from the resistors, to allow a wider experimental investigation on the influence of the heaters on interferometers in different geometric \adde{conditions.


Resistance} of the fabricated heaters was measured, giving an average value of 67~$\Omega$ for the  resistor and 13~$\Omega$ for the plates and connections. The chip was finally mounted on an aluminium base (4.5~cm$\times$6~cm$\times$1~cm) which serves as thermostat, to stabilize the boundary temperature during the device operation.

\subsection{Characterization setup}

\begin{figure*}[t!]
\centering
\includegraphics[width=0.95\textwidth]{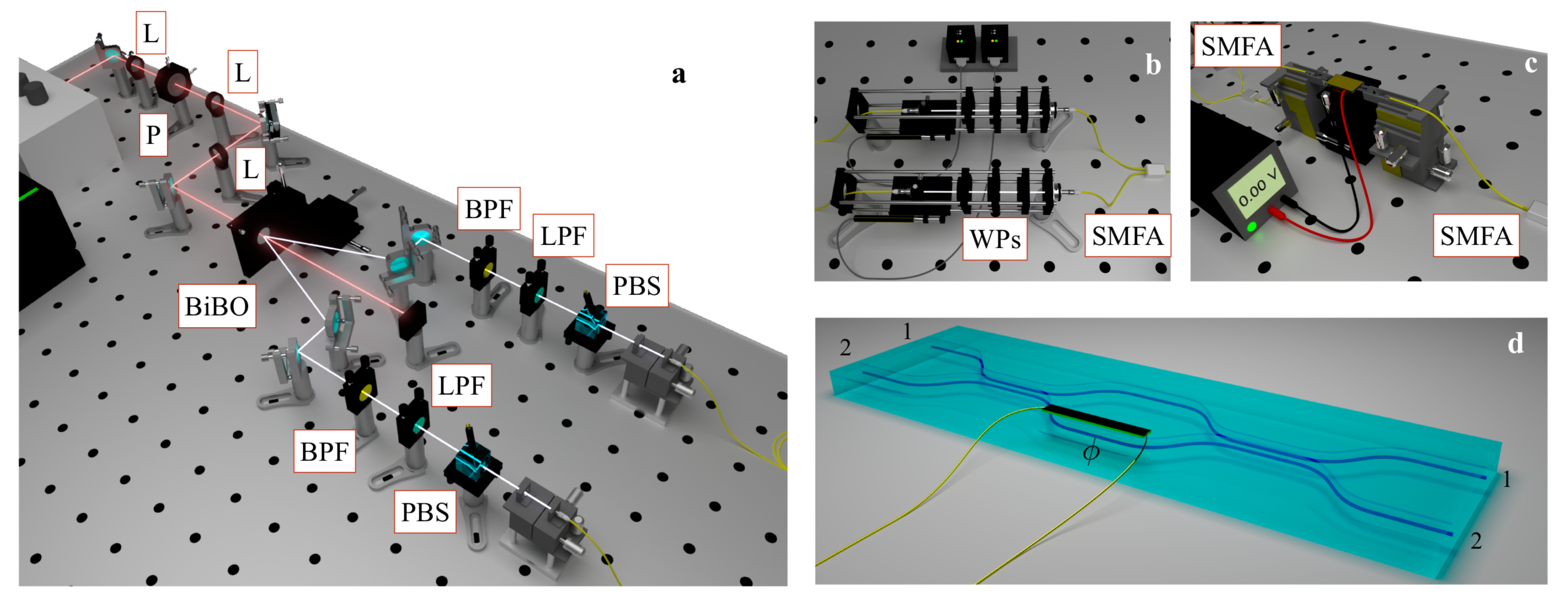}
\caption{ Setup for device characterization with quantum light in the telecom band. ({\bf a}), Schematic of the two-photon source adopted in the experiment: pairs of photons are emitted in two different spatial modes and injected into single-mode fibres. L = Lens \adde{(50~cm focal length)}; P = Pinhole; BiBO = bismuth borate crystal; BPF = Band Pass filter; LPF = Long Pass Filter; PBS = Polarizing Beam Splitter; WPs = Wave Plates for polarization compensation. ({\bf b}), The generated photons are then injected in spatial delay lines for temporal synchronization, propagating through a set of waveplates for polarization compensation. ({\bf c}), Single mode fibre arrays (SMFA) are used to inject the photons in both the interferometer's inputs and to collect photons from both output modes. ({\bf d}),  Schematic of the integrated tunable Mach-Zehnder interferometer. The phase $\phi$ in the interferometer is controlled by tuning the driving voltage on the resistive heating plate, placed on the chip surface.}
\label{fig:setup}
\end{figure*}

The fabricated devices were first characterized with classical light input. To this purpose, light from a tunable laser source (Santec MLS-2000), set at 1550~nm wavelength, was injected in the chosen device input by a standard telecom fibre. The output light from the chip was collected by a 0.65~NA objective. For static measurements, both outputs of the interferometer were simultaneously monitored using two photodiode heads of a Coherent~LabMaster~Ultima power meter, while the driving voltage to the resistors was provided by a TDK-Lambda Zup~6-66 stabilized power supply. Both the optical power meter and the voltage supply were computer controlled and this allowed to completely automatize the voltage scans used to characterize the static thermal response of the interferometers. The same setup was used for the stability measurement. On the contrary, to measure the dynamic step response, the resistor was driven by a function generator (Textronix CFG 280) and an InGaAs photodiode (Thorlabs PDB410C) was used as detector, connected to an oscilloscope (Tektronix DPO 2024B); a single output of the interferometer was monitored for this measurement.

\adde{
For the characterization with quantum light couples of horizontally-polarized identical photons were generated at 1550~nm wavelength, by spontaneous parametric down-conversion (SPDC), in a Type-I bismuth borate (BiBO) biaxial crystal \cite{Bonn12} (see Fig.\ref{fig:setup}a). The crystal was pumped by 160~fs laser pulses at 76~MHz repetition rate, from a Ti:Sapphire oscillator (775~nm wavelength), with the possibility to switch to CW mode. Typical detected count rates were about 16~kHz for singles and 0.5~kHz for coincidences in pulsed regime with an incident power of 600~mW, and about 17 kHz for singles and 2 kHz for coincidences in CW mode with an incident power of 900~mW. A longpass (LPF) and a bandpass (BPF, 8.8~nm bandwidth centred at 1550~nm) interference filters were used in each of the two SPDC modes, to perform a narrow wavelength selection prior to injection into single mode-fibres for spatial selection. A pair of delay lines (see Fig.~\ref{fig:setup}), equipped with a set of waveplates, was used for path synchronization and polarization compensation

Input and output coupling to the chip was carried out by a pair of fibre arrays, each mounted on a micrometric roto-translational stage. 


Detection was performed by a pair of InGaAs/InP single-photon avalanche detectors (ID210 by ID Quantique). Coincidences were counted by an external electronic board, connected to a computer for data analysis. Details of the detectors operation are reported in \suppnote{1}.}

Indistinguishability of the downconverted photons was tested by performing Hong-Ou-Mandel (HOM) interference measurements \cite{Hong87} in a symmetric 50/50 single-mode fibre beam-splitter, both in the pulsed and CW regimes. Raw visibility values were  $\mathcal{V}_{\mathrm{HOM}}^{(1,1)} = 0.967\pm0.002$ in the CW at 800~mW incident power and  $\mathcal{V}_{\mathrm{HOM}}^{(1,1)} = 0.923\pm0.004$ in the pulsed regime at 200~mW incident power \adde{($V^{(1,1)}_{\mathrm{HOM}} = 0.986 \pm 0.002$ and $\mathcal{V}^{(1,1)}_{\mathrm{HOM}} = 0.994 \pm 0.006$ by correcting for the accidental coincidences, in CW and pulsed mode respectively)}, comparable to what has been observed in the literature \cite{Bonn12,Bruno13}. The visibility $\mathcal{V}^{(i,j)}_{\mathrm{HOM}}$ is defined as $\mathcal{V}_{\mathrm{HOM}}^{(i,j)} = (N^{(i,j)}_{\mathrm{clas}}-N^{(i,j)}_{\mathrm{quan}})/N^{(i,j)}_{\mathrm{clas}}$, where $N^{(i,j)}_{\mathrm{clas}}$ is the number of detected coincidences for distinguishable photons and $N^{(i,j)}_{\mathrm{quan}}$ for indistinguishable photons. The superscript $(i,j)$ stands for the number of photons $i$ and $j$ measured on the two output modes of the device.

\section{Results and discussion}

\adde{
\subsection{Device operation} }
\label{sec:invThermal}

\begin{figure}[t!]
\centering
\includegraphics{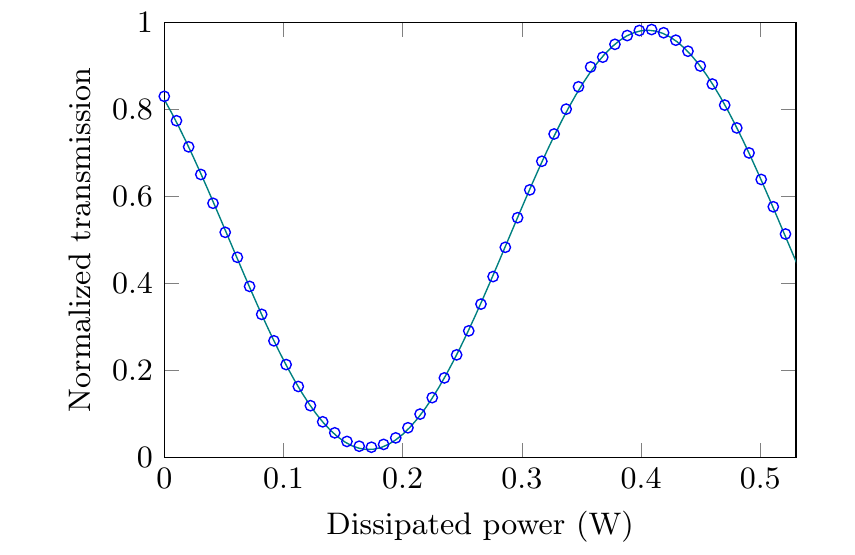}
\caption{Experimental characterization of the interferometer operation with classical light at 1550~nm wavelength.
The transmission of the interferometer in the cross-arm \adde{is plotted} as a function of the thermal power dissipated by the resistor. The best fit of the data with the function reported in Eq. \eqref{eq:MZIresponse} is also shown as a solid line. 
}
\label{fig:charMZI1}
\end{figure}

\adde{
A relative phase delay is induced between the two arms of the interferometer by the temperature increase caused by the resistive heater. Given the linearity of heat equation, which governs the heat diffusion from the resistor in the glass, the temperature variation in a certain point of the glass is proportional to the dissipated power $\mathcal{P}$ (see \suppnote{2}). The temperature-induced refractive index variation in the glass is also linear, at least in the limited temperature operation range that here is considered. Thus, the induced phase difference between the arms $\phi = \varphi_1-\varphi_2$ is linear with $\mathcal{P}$.

Considering a single resistor acting on a single interferometer, for coherent light injected in one input, interference fringes will be observed monitoring the intensity of either one of the two outputs, following:
\begin{equation}
I_{\mathrm{out}} = \frac{I_\mathrm{tot}}{2} \left[ 1 + \mathcal{V} \cos \phi \right]=  \frac{I_\mathrm{tot}}{2} \left[ 1 + \mathcal{V} \cos \left( \Phi_0 + \alpha \mathcal{P} \right) \right]
\label{eq:MZIresponse}
\end{equation}
where $I_\mathrm{tot}$ is the sum of the intensities on the two outputs, $\mathcal{V}$ is the fringe visibility, $\Phi_0$ is a phase term present in the interferometer when no power is applied to the heater and $\alpha$ is a constant which depends on all the geometric, thermal and optical properties of the interferometers. The fringe visibility is influenced by several factors, such as degree of coherence of the input light, splitting ratio of the directional couplers composing the interferometers and differential losses in the two arms. In addition, in case of ohmic resistive heaters, as in our case, one can write $\mathcal{P} = \Delta V^2 / R$ where $\Delta V$ is the imposed voltage on the resistor and $R$ the resistance value.
Note that the phase depends on $\mathcal{P}$, but is independent from the absolute values of the temperatures involved; i.e., thermal drifts of the environment do not affect the interferometer behaviour.

The response of the fabricated interferometers was experimentally characterized as a function of the heat dissipated on different resistors. To this purpose, laser light was fibre-coupled to one input of each interferometer and the optical power from both outputs was measured while varying the voltage across the resistor; this allowed us to fit for each of them the relevant parameters, $\alpha$, $\Phi_0$ and $\mathcal{V}$, in Eq. \eqref{eq:MZIresponse}. Figure~\ref{fig:charMZI1} shows the normalized output intensity, $\overline{I}=I_{\mathrm{out}}/I_\mathrm{tot}$, for the interferometer as a function of the power dissipated by a resistor placed directly above its shallower arm. A best-fit of Eq. \eqref{eq:MZIresponse} yields $\alpha$~=~13.43~rad~W$^{-1}$ and $\Phi_0$~=~0.837~rad, while the fringe visibility is $\mathcal{V} = 0.964 \pm 0.003$. The good agreement between the fitting curve and the experimental points confirms the linearity of the differential phase $\phi$ with the dissipated power, as theoretically discussed above. Although the Mach-Zehnder interferometer is nominally balanced, we observed a value $\Phi_0 \neq 0$. This can be explained by the fact that the two arms of the interferometer are written at different depth in the glass substrate. In fact, small variations in the spherical aberrations of the writing beam are responsible for a slightly different refractive index change induced in the two arms, thus introducing the $\Phi_0$ phase bias.

The experimental calibration previously discussed is necessary to retrieve the precise $\alpha$ and $\Phi_0$ values for a given device; however, it may be also desirable to develop simple models to estimate a priori the thermal response of the interferometers, at least for certain standard circuit geometry. In fact, this would greatly facilitate the design of more complex reconfigurable circuits. We will discuss in the following a simple analytical model that applies to our circuit layout.

Assuming a wire-like heater and neglecting power dissipation from the top surface of the glass (air is an insulator) a logarithmic decay law for the temperature, as a function of the distance from the heater, can be easily worked out (see \suppnote{2}). The latter approximates well the temperature distribution around the resistor, as confirmed by comparison with a numerical simulation of the heat diffusion (see \suppfig{1}).
Within this simplified model, an analytic expression for $\alpha$ in Eq.~\eqref{eq:MZIresponse} can also be derived, given the distances $\rho_1$ and $\rho_2$ of the two interferometer's arm from the heater:
\begin{equation}
\alpha=\frac{2 n_T}{\lambda} \frac{1}{\kappa} \frac{L_\mathrm{arm}}{L_\mathrm{wire}} \ln \frac{\rho_1}{\rho_2}
\label{eq:logModel}
\end{equation}
where $\kappa$ is the thermal conductivity of the glass substrate, $n_T$ is its thermo-optic coefficient, $\lambda$ is the wavelength, $L_\mathrm{arm}$ is the length of the straight segments in the interferometer's arms and $L_\mathrm{wire}$ is the length of the wire-like heater.
The $\alpha$ coefficients were experimentally characterized for different couples of interferometers and resistors, having different ratios of $\rho_1$ over $\rho_2$. Eq.~\eqref{eq:logModel} was found to predict the experimental values with an average accuracy of 10\% (see \suppfig{2}). Thus, this simple model can be used as a robust approximation for designing circuits and estimating the influence of the wire-like resistive heaters on interferometers placed at different distances. 

In case of several resistive heaters (and several interferometers) on the same substrate, cross-talks are possible on each interferometer due to the combined effects of the different heaters.
However, from the linearity of heat equation, one can easily infer the phase $\phi$ induced on a single interferometer as:
\begin{equation}
\phi = \varphi_1-\varphi_2 = \Phi_0 + \sum_i \alpha_i \mathcal{P}_i 
 \label{eq:MZIresponse2}
\end{equation}
where $\mathcal{P}_i$ are the dissipated powers from the different resistors. Note that the interferometer is influenced in principle by all the heaters, but with different constants, depending on the geometric configuration. Of course, $\alpha_i$ for distant heaters is expected to be negligible. However, the influence of each heater to each interferometer could be characterized independently (thus characterizing the various $\alpha_i$) and the combined behaviour could be predicted just by adding the different contributions. This was confirmed experimentally by measuring the output of an interferometer under the action of two heaters, one placed above its shallower arm and one placed above its deeper arm (see \suppfig{3}), and showing that the response can be accurately predicted even in the case of strong cross-talks of different heaters on the same interferometer.

The operation of multiple heaters on the same chip is not a problem also from the heat sinking point of view; in fact, the power that needs to be dissipated will be rather low, thanks to the fact that a 2$\pi$ phase shift in a single interferometer is achieved with only 0.5 W (see Fig.~\ref{fig:charMZI1}). In addition, the circuit reconfiguration, even for a repeated number of times, did not affect the alignment between the chip and the input optical fibres. In fact, the induced temperature variations are in the order of ten degrees Celsius and localized around the heater; thus, they do not induce significant deformation of the whole chip. 

In applications such as quantum optics experiments, where measurements may take from several minutes to several hours or days, operation stability for a reconfigurable circuit is crucial. 
The operation of the interferometer has been monitored for 10 hours under constant voltage driving conditions, imposing a phase of about $\pi/2$ where the sensitivity to variations in voltage is maximum (see \suppfig{4}): observed phase fluctuations have been smaller than 0.01~rad with no evidence of drifts.
Finally, the response of the device to a voltage step has been also characterized (see \suppfig{5}), showing a rise time of $\sim$0.9~s (10\% to 90\% of the variation). This means that the circuit is completely settled on a new configuration within a few seconds.
}

\subsection{Two-photon interference}

The overall action $U^{\mathrm{theo}}$ of an ideal Mach-Zehnder interferometer on the input-output modes $b_{i}^{\dag} = \sum_{j} U^{\mathrm{theo}}_{i,j} a_{j}^{\dag}$,  being $a^{\dag}_{i}$ and $b^{\dag}_{i}$ the field operators for the input and output modes respectively, is described by:
\begin{equation}
U^{\mathrm{theo}} = \begin{pmatrix} \sin \phi & \cos \phi\\ \cos \phi & - \sin \phi \end{pmatrix}
\end{equation}
where a global phase factor $\imath e^{-\imath \phi/2}$ has been removed.

\begin{figure}[t]
\centering
\includegraphics[width=8cm]{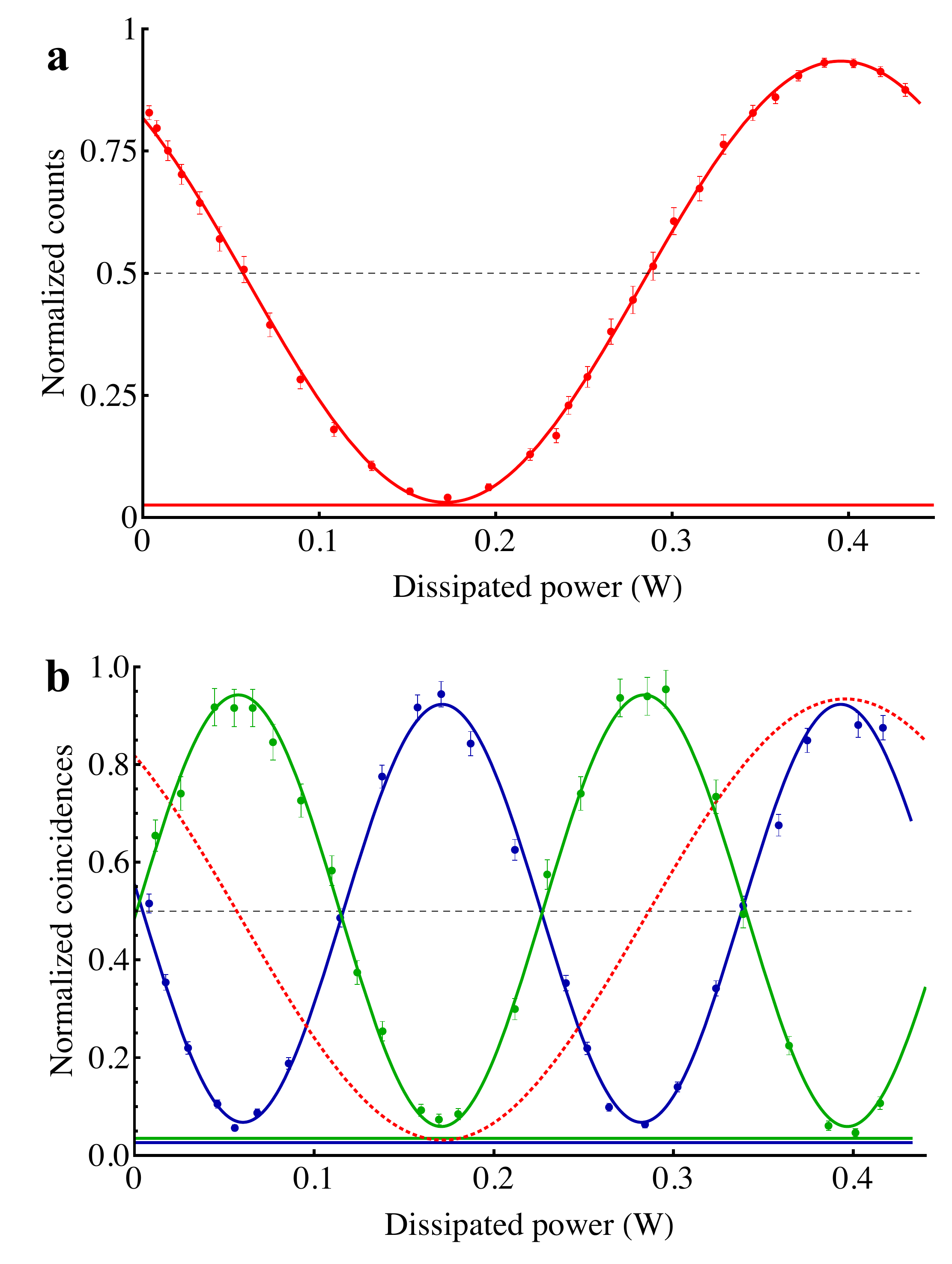}
\caption{({\bf a}), Experimental measurement of a single-photon fringe pattern for input mode $1$ and output mode $2$, as a function of the dissipated power. Points: normalized counts. Solid red curve: best fit of the measured data with a sinusoidal function. Horizontal red line: estimated contribution of accidental coincidences. ({\bf b}), Experimental fringe pattern for a two-photon input $\vert 1,1 \rangle$ as a function of the dissipated power. Blue points: normalized coincidences of the output $(1,1)$, with one photon for each output port; statistics of the recorded data was $\sim 1200$ coincidences for the fringe maximum in 120s. Green points: normalized coincidences of the output $(0,2)$, with two photons on output port 2; statistics of the recorded data was $\sim 600$ coincidences for the fringe maximum in 180s. Both fringes show twice the periodicity of single-photon ones (dashed red curve). Solid lines: best fit of the measured data with a sinusoidal function. Horizontal blue and green lines: estimated contribution of accidental coincidences. The error bars are given by Poissonian statistics.}
\label{twophoton}
\end{figure}
We first measured single-photon fringe patterns, by injecting an heralded single photon in input port $1$ and by measuring the triggered count rate at output port $2$ (see Fig. \ref{twophoton}a). The measured raw visibility was $\mathcal{V}=0.930 \pm 0.006$ ($\mathcal{V}=0.981 \pm 0.007$ correcting for the accidental coincidences), compatible with the one performed with classical light. We then recorded the interference fringes obtained by injecting two photons into the device in a $\vert 1,1 \rangle$  state, where $\vert i,j \rangle$ stands for $i$ ($j$) photons on input port $1$ ($2$). In this case, after the first beam-splitter the state evolves into a two-photon N00N state, $\propto (\vert 2,0 \rangle - \vert 0,2 \rangle)$. The action of the phase shift changes the state into $\propto (\vert 2,0 \rangle - e^{-\imath 2 \phi} \vert 0,2 \rangle)$. The fringe pattern recorded in the two-fold coincidences after the recombination in the second beam-splitter, corresponding to the measurement of the $(1,1)$ output state, lead as expected to a \adde{halved} period with respect to the single-photon \adde{case}. The experimental fringe pattern is shown in Fig. \ref{twophoton}b. We measured an experimental raw visibility $\mathcal{V}^{(1,1)}_{\mathrm{N00N}} = 0.864 \pm 0.007$ ($\mathcal{V}^{(1,1)}_{\mathrm{N00N}} = 0.913 \pm 0.006$ by subtracting the accidental coincidences), showing the correct operation of the integrated interferometer. The visibility $\mathcal{V}^{(i,j)}_{\mathrm{N00N}}$ is defined as $\mathcal{V}^{(i,j)}_{\mathrm{N00N}}=(N_{\mathrm{max}}^{(i,j)} - N_{\mathrm{min}}^{(i,j)})/(N_{\mathrm{max}}^{(i,j)}+N_{\mathrm{min}}^{(i,j)})$, where the subscripts refer to the maximum and the minimum number of coincidences measured scanning the phase $\phi$.

We then measured the events in which the two photons emerge from the same output port $2$, by connecting it to a symmetric 50/50 fiber beam-splitter and by collecting two-fold coincidences between the output of the beam-splitter. The results are shown in Fig. \ref{twophoton}b: the $(0,2)$ output contribution exhibits the same super-resolved two-fold pattern with opposite relative phase with respect to the $(1,1)$ contribution. The measured visibility was $\mathcal{V}^{(0,2)}_{\mathrm{N00N}} = 0.882 \pm 0.008$, which increases up to $\mathcal{V}^{(0,2)}_{\mathrm{N00N}} = 0.949 \pm 0.007$ by subtracting the contribution from accidental coincidences.

To assess the non-classicality of the observed super-resolved two-photon interference fringes, we applied a recently proposed criterion \cite{Afek10}, which compares the visibility of the pattern with what can be achieved with classical light. While it is possible to obtain fringes oscillating as $2 \phi$ with perfect visibility $\mathcal{V}^{(1,1)}=1$ with appropriately tailored classical light measuring the $(1,1)$ output contribution, the maximum achievable visibility for the $(0,2)$ output contribution is bounded to $\mathcal{V}_{\mathrm{cl}}^{(0,2)} < 1/3$. The raw visibility $\mathcal{V}^{(0,2)}_{\mathrm{N00N}}$ measured with two-photon input exceeds the classical bound by more than $68$ standard deviations ($88$ for the corrected value $\mathcal{V}^{(0,2)}_{\mathrm{N00N}}$), thus showing the quantumness of the observed effect.

\begin{figure}[t]
\centering
\includegraphics[width=8cm]{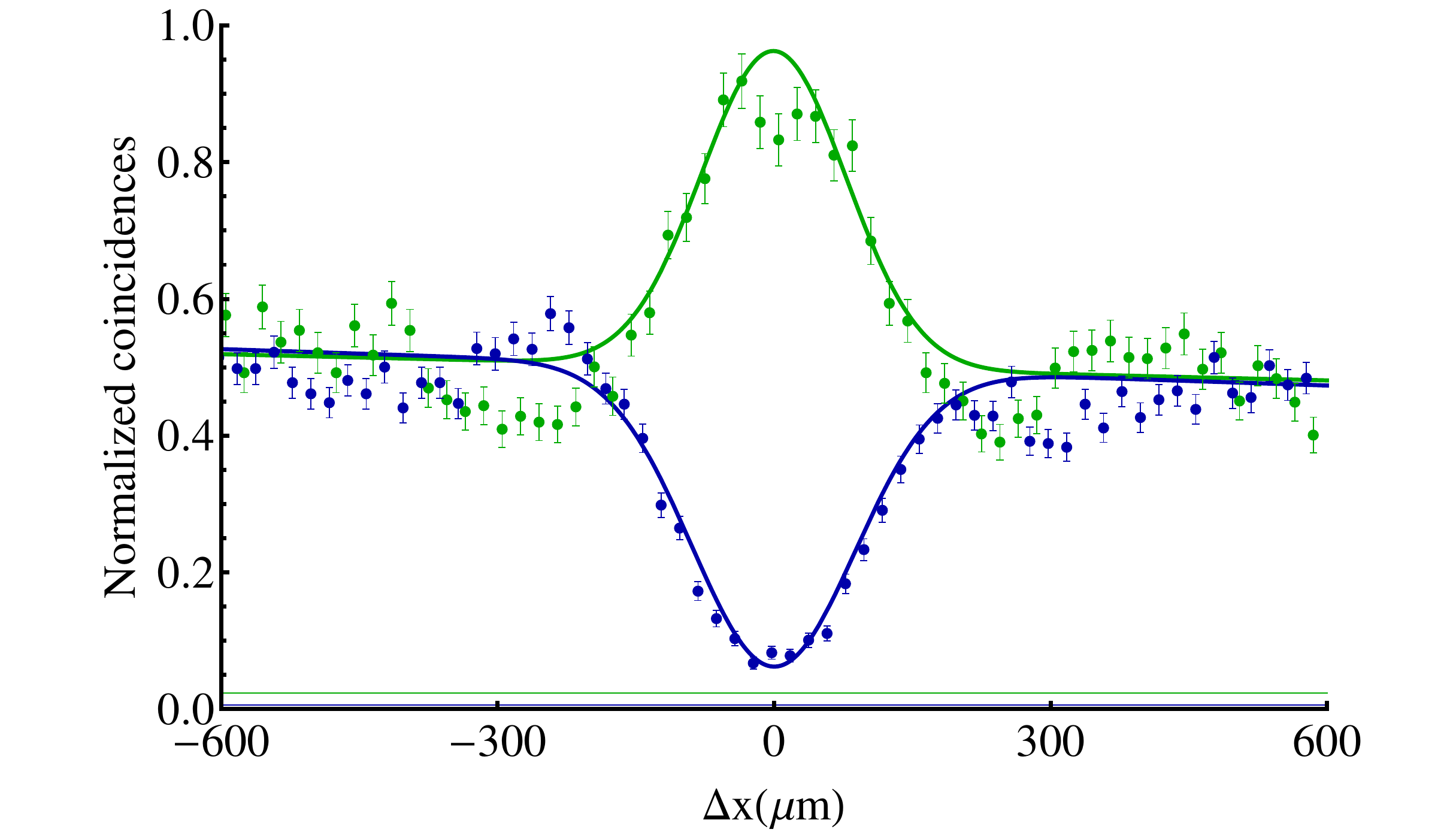}
\caption{Hong-Ou-Mandel interference experiment for input state $\vert 1,1 \rangle$ at $\Delta V \simeq 1.97$ V ($\phi \sim \pi/2$). Blue Points: coincidences for output $(1,1)$. Green points: coincidences of output $(0,2)$. Solid lines: best fit of the measured data. Horizontal lines: levels for the contribution of accidental coincidences. The error bars are given by Poissonian statistics.}
\end{figure}

We finally performed a Hong-Ou-Mandel interference experiment for a voltage set to $\Delta V\simeq 1.97$ V, corresponding to $\phi \simeq \pi/2$, where the interferometer behaves as a 50/50 beam-splitter. We measured both the (1,1) and (0,2) output contributions as a function of the optical delay between the photons, leading to experimental raw visibilities of $\mathcal{V}^{(1,1)}_{\mathrm{HOM}} = 0.876 \pm 0.010$ and $\mathcal{V}^{(0,2)}_{\mathrm{HOM}} = - 0.925 \pm 0.039$ respectively ($V^{(1,1)}_{\mathrm{HOM}} = 0.886 \pm 0.010$ and $\mathcal{V}^{(0,2)}_{\mathrm{HOM}} = - 0.976 \pm 0.041$ by correcting for the accidental coincidences). \adde{We attribute the reduced visibility to a voltage set slightly different from the one corresponding to a 50/50 beam splitter.}

\subsection{Tomography of the interferometer}

\begin{figure*}[ptb]
\centering
\includegraphics[width=0.8\textwidth]{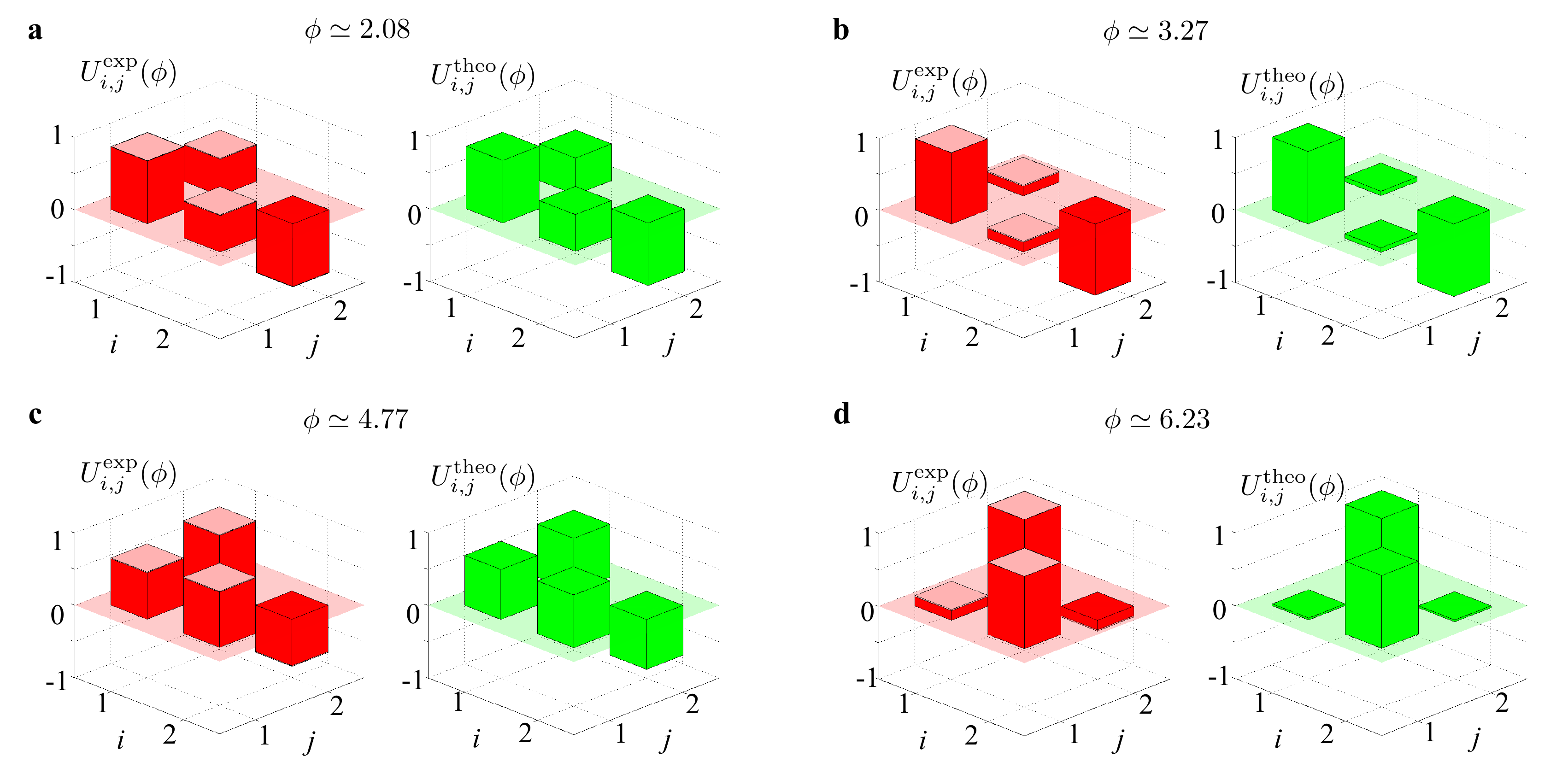}
\caption{Tomography of the interferometer for different values of the phase $\phi$. ({\bf a}), Tomography for $\phi \simeq 2.08$. ({\bf b}), Tomography for $\phi \sim \pi$, where the interferometer behaves as the identity with an additional (-1) phase factor on mode $2$. ({\bf c}), Tomography for $\phi \sim 3 \pi/2$: the interferometer behaves as a symmetric 50/50 beam splitter. ({\bf d}), Tomography for $\phi \sim 2 \pi$: the interferometer behaves as a SWAP gate, exchanging modes $1$ and $2$. For each panel: red bars correspond to the reconstructed unitary $U^{\mathrm{exp}}$ while green bars correspond to the ideal unitary $U^{\mathrm{theo}}$. Light red bars correspond to the error in the reconstruction process.}
\label{fig:Tomography}
\end{figure*}

To fully characterize the action of the interferometer, we performed the tomography of the corresponding unitary transformation for different values of the phases $\phi$. Tomography of linear transformation can be performed by applying a recently proposed method \cite{Peru11,Lain12} based on single-photon probability measurements, or equivalently intensity measurements with classical light, and on two-photon interference Hong-Ou-Mandel experiments. The action of the unitary linear transformations can then be retrieved by applying a suitable analytical \cite{Lain12}, numerical \cite{Peru11} or hybrid \cite{Cres13} algorithm.

In Fig. \ref{fig:Tomography} we report the results of the tomographic characterization of the device for $4$ different relevant values of the phase. We observe a very good agreement between the expected behaviour and the measured one.
The average gate fidelity $F= 1/2 \vert \mathrm{Tr}[{(U^{\mathrm{theo}})}^{\dag} U^{\mathrm{exp}}] \vert$, calculated for $12$ different values of $\phi$, is $\overline{F} = 0.998 \pm 0.001$, thus showing the quality of the fabrication process.

\section{Conclusions}
We have demonstrated the fabrication by femtosecond lasers of an integrated Mach-Zehnder interferometer at an operating wavelength of 1550~nm, which can be dynamically reconfigured by thermal shifters. Measurements with classical light, single-photons and two-photon states confirm the correct functioning of the device, including the observation of quantum interference fringes corresponding to a two-photon N00N state. Full characterization of the thermal response of the interferometric circuit has also been performed, showing good agreement with what expected from fabrication design. This device represents a possible building block of future reconfigurable quantum circuits in the telecom band, paving the way to the use of integrated photonic devices to implement large quantum networks.

{While this work was being written, reconfigurable operation of a femtosecond laser written interferometer at 800~nm wavelength was reported \cite{Chab14}.}

\section{Acknowledgments}

This work was supported by ERC-Starting Grant 3D-QUEST (3D-Quantum Integrated Optical Simulation; grant agreement no. 307783, http://www.3dquest.eu), and by Marie Curie Initial Training Network PICQUE (Photonic Integrated Compound Quantum Encoding, grant agreement no. 608062, funding Programme: FP7-PEOPLE-2013-ITN, http://www.picque.eu).

\cleardoublepage

\section*{Supplementary Note 1 - Details of the single-photon detectors operating mode}

To achieve best performance, the SPADs (ID210 by ID Quantique) were used in two different operating modes: one was employed in free-running mode, while the other operated in external gating mode, triggered by the output of the first detector, thus significantly reducing the dark count rate (DCR). In free-running mode, the SPADs' efficiency was set to 10\% and the best SNR ratio was attained by setting the dead time at 30 $\mu$s giving a 2.5~kHz DCR. The gated detector was set at 100~$\mu$s dead time and 25\% efficiency, giving rise to a DCR of less than 2~Hz. In order to compensate the delay of $\sim 70$~ns induced by the detection and the output pulse emission in the first detector, a 15~m single mode fibre together with a tunable internal gate delay were employed in the gated detector's path.

\section*{Supplementary Note 2 - Modelling the thermal diffusion}

For a general arbitrary configuration of heat sources and materials, heat diffusion is governed by:
\begin{suppEq}
q(\vec{r},t) = - \kappa(\vec{r}) \nabla^2 T (\vec{r},t) + c(\vec{r}) \frac{\partial T(\vec{r},t)}{\partial t}
\label{eq:calore}
\end{suppEq}
where $T(\vec{r},t)$ gives the temperature distribution, $q(\vec{r},t)$ is the thermal power generated per unit volume, $\kappa(\vec{r})$ is the thermal conductivity of the material and $c(\vec{r})$ is the heat capacity per unit volume.

Let us analyse the static case, in which we drop the time dependency in Eq.~\eqref{eq:calore}. In case of heat generated by a single resistor with a given geometry, dissipating a thermal power $\mathcal{P}$, linearity of Eq.~\eqref{eq:calore} implies a solution of the form:
\begin{suppEq}
T(\vec{r}) = T_0 + \mathcal{P} \, g(\vec{r}) 
\end{suppEq}
where $T_0$ is a base temperature which depends on boundary conditions (it can be assumed as the environment temperature) and $g(\vec{r})$ is a proper spatial function, which depends on the geometry of the resistor.
For limited temperature modulations, the dependence of the material refractive index on the temperature can be assumed linear and expressed by means of the thermo-optic coefficient of the material $n_T$ as $n(\vec{r}) = n_0(\vec{r}) + n_T \left( T(\vec{r}) - T_0 \right)$. It is then easy to show that the phase delay accumulated by light propagating in an arbitrary waveguide segment is a linear function of the dissipated power $\mathcal{P}$ as follows:
\begin{suppEq}
\varphi(P) = \varphi_0 + \frac{2 \pi}{\lambda} \mathcal{P} n_T \int_L g(\vec{r}) dl
\end{suppEq}
with $\varphi_0 = \frac{2 \pi}{\lambda} \int_L n_0 (\vec{r}) dl $.
Therefore, considering a Mach-Zehnder interferometer, the difference between the phase delays $\varphi_1$ and $\varphi_2$ accumulated in the two arms can be written as:
\begin{suppEq}
\phi = \varphi_1 - \varphi_2 = \Phi_0 + \alpha \mathcal{P}
\end{suppEq}
being $\Phi_0$ an offset phase, present in the interferometer when no power is applied to the heater and 
\begin{suppEq}
\alpha = \frac{2 \pi}{\lambda} n_T \left( \int_{L_1} g(\vec{r}) dl - \int_{L_2} g(\vec{r}) dl \right)
\end{suppEq}
The integrals $\int_{L_1} g(\vec{r}) dl$ and $\int_{L_2} g(\vec{r}) dl$ are performed respectively along the paths of the first and second arm of the interferometer.

We will now discuss a simple analytical model that allows to estimate approximately the value of $\alpha$ for the device layout adopted in our experiments.
We have chosen to use resistive heaters much longer than wide, that are assimilable to wires. In a reference system with polar coordinates $\rho, \theta$ centred on one wire, the expression of the heat source is $q=\delta(\rho)$. If such wire is surrounded by a totally isotropic medium, the thermal diffusion follows a cylindrical symmetry and the solution to Eq.~\eqref{eq:calore} would be:
\begin{suppEq}
T = C_1 - C_2 \ln \rho
\label{eq:Tiso}
\end{suppEq}
with proper integration constants $C_1, C_2$ dependent on the boundary conditions. {In this configuration the heat diffuses radially; if we consider, for instance, a virtual horizontal surface, the temperature gradient is always parallel to it and there is no net heat flux through such surface.
Here, the wire-like heater} is deposited on a surface which acts as interface between a strong insulator (air, with thermal conductivity $\kappa_{\mathrm{air}} \simeq 0.026$~Wm$^{-1}$K$^{-1}$) and a thermal conductor (glass, with thermal conductivity $\kappa_{\mathrm{glass}} \simeq 1$~Wm$^{-1}$K$^{-1} \; \gg \kappa_{\mathrm{air}}$). The heat dissipated in air is negligible with respect to the heat diffused in the glass. Thus, we can approximate the temperature distribution in glass with that calculated in the isotropic case (Eq.~\eqref{eq:Tiso}), but considering all the heat generated by the wire as dissipated on the glass side.  In addition, we will consider that the temperature-induced phase change is active only on the straight segments of the interferometer's arms, placed respectively at distance $\rho_1$ and $\rho_2$ from the hot wire.
Within these approximations, it is easy to derive:
\begin{suppEq}
\alpha=\frac{2 n_T}{\lambda} \frac{1}{\kappa} \frac{L_\mathrm{arm}}{L_\mathrm{wire}} \ln \frac{\rho_1}{\rho_2}
\label{eq:SlogModel}
\end{suppEq}
where $L_\mathrm{arm}$ is the length of the straight segments in the interferometer's arms and $L_\mathrm{wire}$ is the length of the wire-like heater.

\renewcommand{\figurename}{Supplementary Figure S -}
\setcounter{figure}{0} 

\begin{figure*}[p]
\centering
\includegraphics{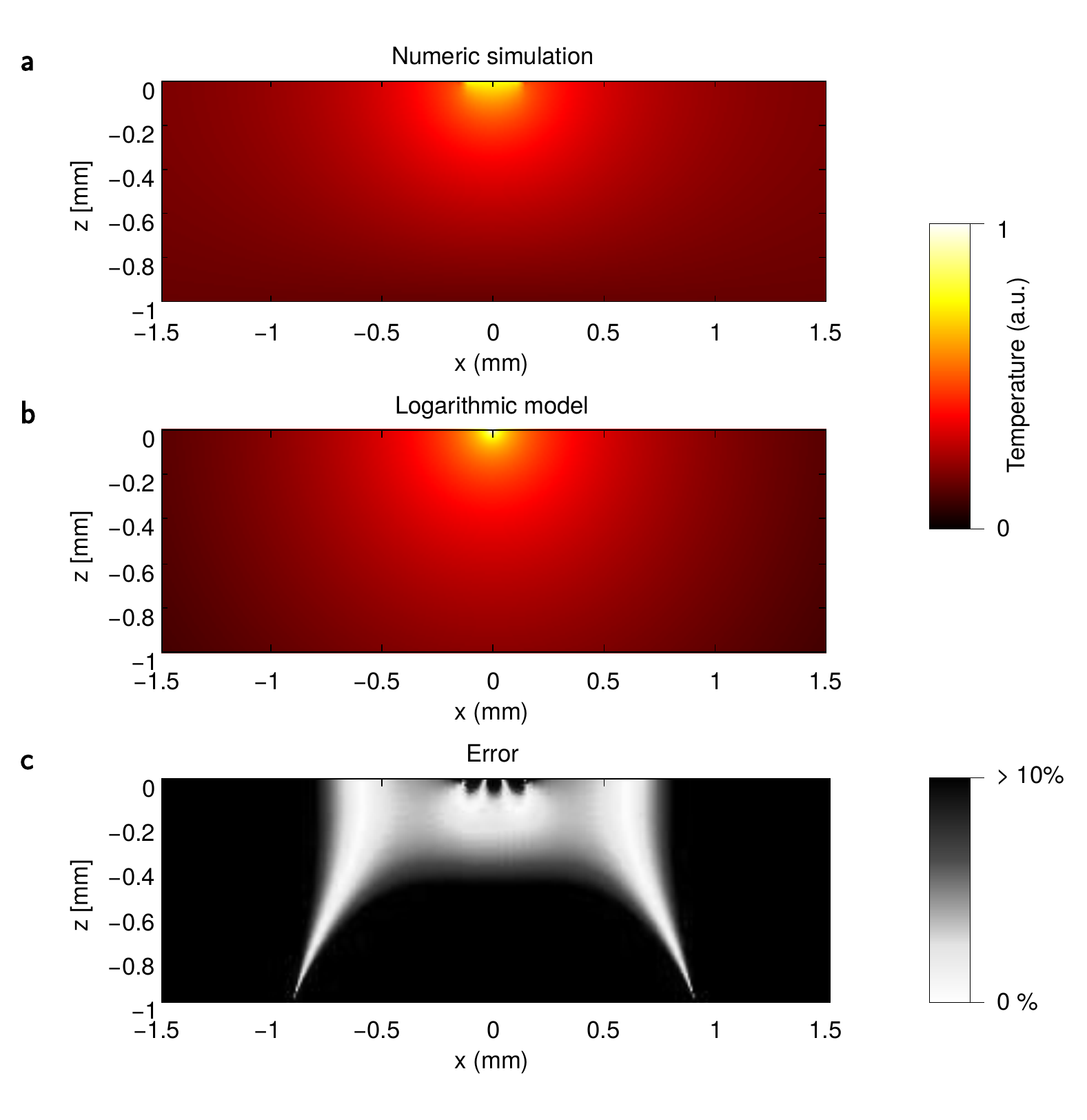}
\caption{(\textbf{a}) Numeric simulation of the temperature distribution inside the glass, considering an infinite-length, 300~$\mu m$ wide, heater, placed on the top surface. The heater and the base of the glass are assumed at uniform (different) temperatures. (\textbf{b}) Temperature distribution for a wire-like heater, assuming a logarithmic decaying law ($T = C_1 - C_2 \log \rho$) with the distance from the heater. Constant $C_2$ is chosen so that the dissipated power through the glass is the same as in the case of the numeric simulation. Constant $C_1$ is just an additive term which has no influence on the shape of the distribution and is chosen to fit the values of the numeric simulation. Temperature distributions in (a) and (b) are normalized to the same color scale (reported). (\textbf{c}) Absolute values of the relative difference between temperature values in the graphs (a) and (b).  White corresponds to null difference, black to difference larger than 10\%. The simple logarithmic model appears to predict quite accurately the temperature distribution in a region $\sim$1.5~mm wide and $\sim$0.5~mm deep around the heater.}
\end{figure*}

\begin{figure*}[p]
\centering
\begin{tabular}{cc}
{\bfseries\sffamily \large a} & \\
& \includegraphics{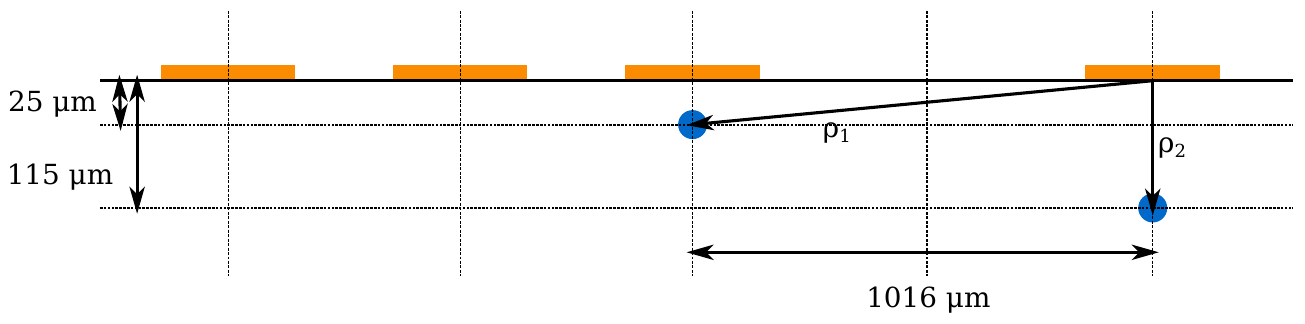}
\\ 
&
\\
{\bfseries\sffamily \large b} & \\
& \begin{tikzpicture}
	\begin{axis}[
		xmin=0,
		ymin=0,
		xmax=2,
		ymax=15,
		height=6cm,
		width=7cm,
		at={(0cm,-0.5cm)},
		anchor=north,		
		xlabel={$|\ln \frac{\rho_1}{\rho_2}|$},
		ylabel=$|\alpha|$ (rad~W$^{-1}$),
		]
		
	\addplot [only marks, color = blue] coordinates {
		(	1.6117180636	,	13.68	)
		(	0.9463273039	,	9.46	)
		(	0.3015929566	,	2.36	)
		(	0.477828939	,	4.68	)
		(	0.4663273006	,	5.46	)
		(	1.6117180636	,	13.1	)
		(	0.2982984979	,	2.32	)
		(	1.6117180636	,	13.16	)
		(	0.9463273039	,	8.64	)
		(	1.6117180636	,	12.5	)
		(	1.6117180636	,	13.43	)
	};

	\addplot+[domain=0:2,mark=none,samples=3, color=black, line width = 0.7pt]{ 8.60215*x};

	\addplot+[domain=0:3,mark=none,samples=3, color=black, line width = 0.3pt, dashed]{ 8.60215*1.1*x};
	\addplot+[domain=0:3,mark=none, samples=3, color=black, line width = 0.3pt, dashed]{ 8.60215*0.9*x};

	\end{axis}
\end{tikzpicture}
\end{tabular}

\caption{(\textbf{a}) Conceptual picture of the geometry of the fabricated device layout: each interferometer has two arms (waveguides represented as blue circles in the picture), one shallower and one deeper. On the glass surface several heaters are realized (orange rectangles in the picture), placed at different distances $\rho_1$ and $\rho_2$ with respect to the two waveguides. (\textbf{b}) Experimental values for the modulus of the $\alpha$ coefficients, measured for interferometers and thermal shifters with different distances between the resistor and the two interferometer arms; the values are plotted as a function of $ \vert \ln \frac{\rho_1}{\rho_2} \vert$. The expected trend is also reported (solid line), which is obtained by substituting in Eq.~\eqref{eq:SlogModel} the values $\lambda$=1.55~$\mu$m, $n_T$~=~1$\times 10^{-5}$~K$^{-1}$, $L_\mathrm{arm}$~=~12~mm, $L_\mathrm{wire}$~=~20~mm, $\kappa$~=~0.9~W~m$^{-1}$~K$^{-1}$ found in the literature. Dashed lines delimit a $\pm$10\% region around the expected trend line; most experimental points fall within this region.}

\end{figure*}
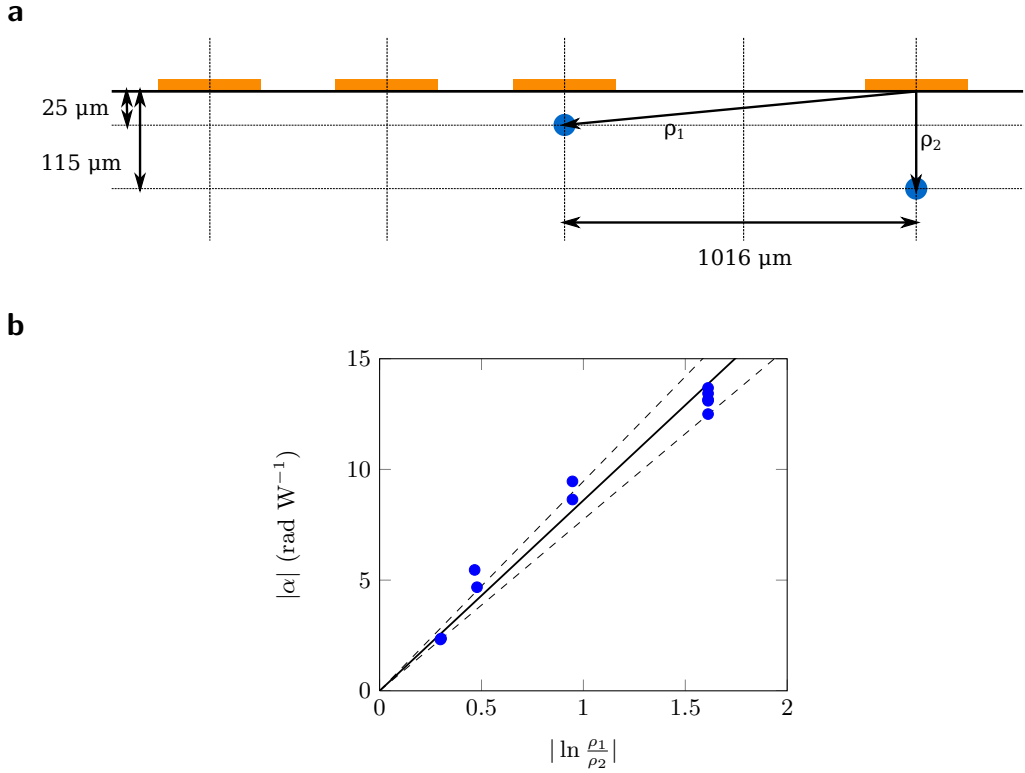

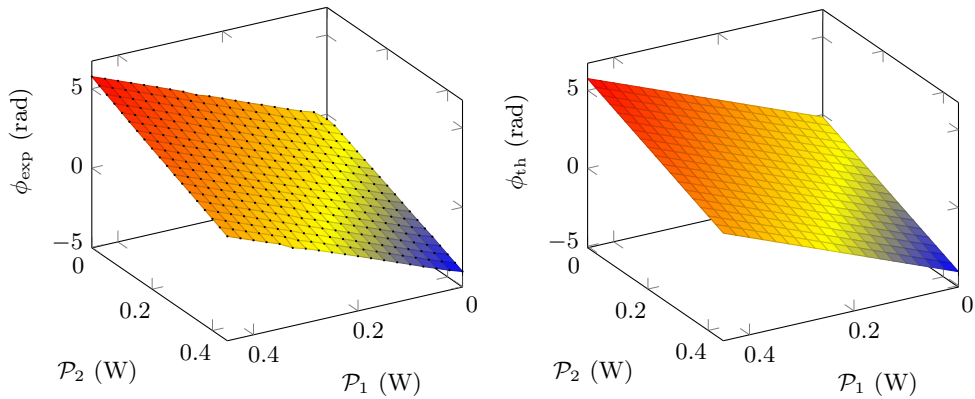
\begin{figure*}[p]
\centering
\begin{tabular}{cc}
	\begin{tikzpicture}[baseline]
		\begin{axis}[name=plot1, width=6.5cm,height=6cm, 
		 view/h=150, xlabel=$\mathcal{P}_1$ (W), ylabel=$\mathcal{P}_2$ (W), zlabel=$\phi_\mathrm{exp}$ (rad)]
			\addplot3[surf, mesh/ordering=x varies, mark = *, mark size = 0.2pt] table {superficieFase.dat};
		\end{axis}
	\end{tikzpicture}
&
	\begin{tikzpicture}[baseline]
		\begin{axis}[name=plot2, width=6.5cm,height=6cm,
		view/h=150, xlabel=$\mathcal{P}_1$ (W), ylabel=$\mathcal{P}_2$ (W), zlabel=$\phi_\mathrm{th}$ (rad)]
			\addplot3[surf, domain = 0:0.45, y domain=0:0.45, samples = 19] {-8.77*y+13.16*x-0.165};
		\end{axis}
	\end{tikzpicture}
\end{tabular}

\caption{An interferometer is characterized under the simultaneous action of two different heaters. The graph on the left reports the experimentally retrieved phase $\phi_\mathrm{exp}$ for 361 different couples of values of dissipated powers on the two heaters. It is noted that the measured values lie on a plane, thus confirming the possibility to express the interferometer's response as a linear superposition of the effects of each resistor. The graph on the right plots, for comparison, the ideal phase $\phi_\mathrm{th} = \Phi_0 + \alpha_1 \mathcal{P}_1 + \alpha_2 \mathcal{P}_2$ predicted from the values $\alpha_1$~=~13.16~rad~W$^{-1}$, $\alpha_2$~=~-8.77~rad~W$^{-1}$ and $\Phi_0$~=~-0.17~rad, characterized previously on the same interferometer by acting on one resistor at once. Standard error between the two graphs is 0.11~rad.
}
\label{fig:charMZI2}
\end{figure*}

\begin{figure*}[p]
\centering
\begin{tikzpicture}
	\begin{axis}[
		xmin=0,
		ymin=1.5,
		xmax=10,
		ymax=1.75,
		height=4cm,
		width=10cm,
		anchor=north,		
		xlabel={Time (h)},
		ylabel={$\phi$ (rad)},
	]
	\addplot [no markers, color = black] table {stabilita.csv};		
	\end{axis}
\end{tikzpicture}

\caption{Phase-stability measurement under constant driving voltage of the Mach-Zehnder interferometer.}

\end{figure*}
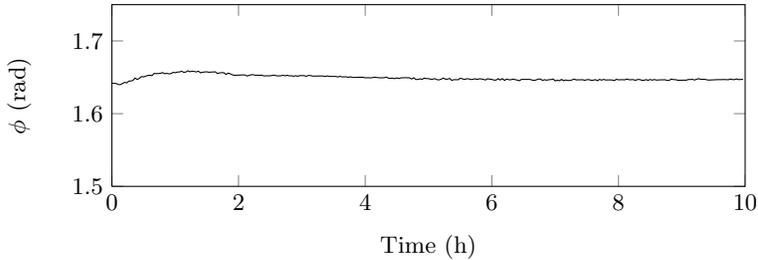

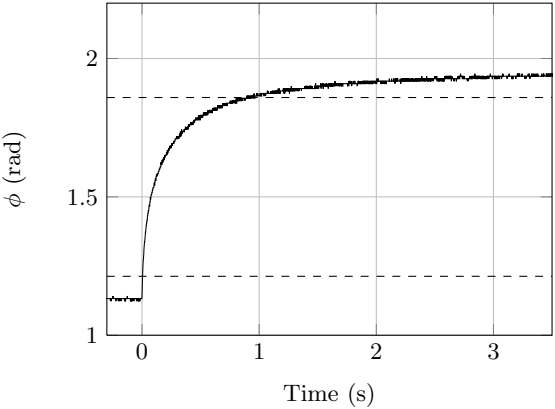
\begin{figure*}[p]
\centering
\begin{tikzpicture}
	\begin{axis}[
		xmin=-0.3,
		ymin=1,
		xmax=3.5,
		ymax=2.2,
		height=6cm,
		width=7.5cm,
		anchor=north,		
		xlabel={Time (s)},
		ylabel={$\phi$ (rad)},
		grid=major,
	]
	\addplot [no markers, color = black] file {riseTime.csv};		
	\addplot[dashed, black] coordinates {(-0.3,1.213) (3.5, 1.213)};
	\addplot[dashed, black] coordinates {(-0.3,1.859) (3.5, 1.859)};
	\end{axis}
\end{tikzpicture}

\caption{Measured time response of the Mach-Zehnder interferometer to an instantaneous voltage step on the heater, from 1.5~V to 2.5~V. Coherent light is coupled to one input and phase $\phi$ is retrieved by measuring the transmitted light power with a fast photodiode. Rise time (10\% to 90\%) is $\sim$0.9~s.}

\end{figure*}

%
%
%

\end{document}